# Foucault-Wheatstone device in the deep-sea

by Hans van Haren

NIOZ Royal Netherlands Institute for Sea Research, P.O. Box 59, 1790 AB Den Burg, the Netherlands.
e-mail: hans.van.haren@nioz.nl

**ABSTRACT**

A pressure sensor, located for 4 months in the middle of a 1275-m long taut deep-sea mooring in 2380 m water-depth above a seamount with sub-surface top-buoys and seafloor anchor-weight, demonstrates deterministic spectral peaks at sub-harmonics of the local near-inertial frequency. None of these frequencies can be associated with oceanographic motions. No corresponding peaks are found in spectra of other observables like water-flow (differences), temperature, and pressure in the top-buoys of the mooring. The mid-cable pressure sensor was mounted on a nearly 1-kN weighing non-swiveled frame. Its data seem to reflect a resonant mechanical oscillation of the tensioned mooring cable under repeated short-scale Strouhal vibrations induced by water-flow. Equivalent vertical motions of 0.5-m amplitudes are found dominant at frequency $f^*/4$, with monotonically decreasing peaks at $f^*/2$, $3f^*/4$ and $f^* = 1.083f$, where $f$ denotes the local inertial frequency of Earth rotation. The observations provide, physically, proof of the Earth rotation and, oceanographically, of the relative vortex-rotation $\xi = 0.083f$ induced by (sub-)mesoscale eddies associated with a nearby seamount, so that effective near-inertial frequency $f^* = f + \xi$. As instrumentation only shows mid-cable resonant signals it is not considered as a Foucault pendulum, but more likely an Earth-rotation equivalent of the Wheatstone device with the mooring cable acting as elastic springs. While the $f^*/2$ motion reflects the Foucault principle of linear orbits fixed in an inertial frame of reference, the dominant sub-harmonic $f^*/4$ suggests resemblance with the precession of a spherical oscillator at half the angular speed of rotation.

# I. INTRODUCTION

The rotation of the Earth has been subject of many studies ranging from astronomy to geophysical flows. Its importance for life cannot be underestimated, as it governs the major displacements of weather systems in the atmosphere (together with the effects of the sun). Likewise in the ocean, the entire large-scale circulation is driven by the Earth rotation (and by wind i.e. the sun). Considering this importance it is not surprising that physicists sought proof of its existence, a cause that was tackled after careful observations by Foucault and the presentation of his pendulum (Foucault, 1851). The pendulum was seen by the gathered crowd to slowly rotate its path of oscillation at a rate of twice the local inertial period, i.e. a period of 24 hours divided by the sine of the local latitude. The principle of the pendulum evidencing the Earth rotation is based on the exchange of angular momentum via the Coriolis-effect (Coriolis, 1835) between a relatively rapidly oscillating device, the pendulum, in a slowly rotating frame of reference, the Earth. Improvements for a properly working device have been given (Kamerlingh Onnes, 1879; Schulz-DuBois, 1970). Theoretically, the Foucault pendulum oscillates in a single plane that remains fixed in an inertial frame. In practice however, effects such as friction cause the oscillation to deteriorate and additional power like transmitted via a magnet is needed to maintain the pendulum-oscillation (Sommeria, 2017).

Besides pendulums, other devices have been constructed to demonstrate Foucault's principle. A few months after the presentation in the "Observatoire de Paris" in 1851, Wheatstone introduced a table-top version, to demonstrate the effect of a turn-table rather than the Earth rotation, by the suspension of a spring between the rotating table and a vertical arc (Wheatstone, 1854). Equivalent versions include a placing of a weight as bob in the middle of the spring, actually by placing the weight between two springs (Teunissen, 2022). The ensemble of springs and weight may be put vertically under a rectangular support at some distance off the rotational axis. No Earth-rotation equivalent is known for such Foucault-Wheatstone devices to the author.

In the ocean, an important means to study flow variables is using the Eulerian principle of fixing instrumentation in space and registering the passing flow at regular intervals in time. Records of variables can be registered in remote deep-sea areas by means of self-contained instrumentation with power and data storage in water-tight and pressure-resistant containers that remain on site for months. Although it is attractive to have the instruments in a 'mooring' suspended on a cable between an anchor-weight at the seafloor and a floatation device or buoy at the surface, e.g., for data-transmission via satellite, such a configuration provides inferior records. This is because the surface-wave and water-flow actions constantly move the mooring cable via resistance-drag as the mooring cannot be tautly tensioned that would be driven to the point of cable-breaking by the wave-action. An improved set-up is using underwater buoys and acoustic release devices for recovery. In an optimization of sufficient buoyancy and minimization of drag using thinner cables and smaller instruments a near-vertical Eulerian mooring can be established with cable tensions O(1-10 kN). At the Netherlands Institute for Sea Research NIOZ, mechanical engineering efforts have resulted in taut-wire moorings that deflect less than O(0.1 m) vertically and O(10 m) horizontally over O(1000 m) lengths under deep-sea flow conditions.

The effort for good sub-surface mooring design (Dewey, 1999) results in improved studies of ocean-interior waves, for which the Earth rotation plays an important role too. The lowest frequency of 'internal waves' supported by typical open-ocean stable vertical density stratification is the inertial frequency or Coriolis parameter $f = 2\Omega\sin\varphi$, with $\Omega$ denoting the Earth's angular velocity and $\varphi$ the latitude. Deep-sea motions at this frequency are generated, e.g., as transients after the passage of a storm. They follow a near-circular path and have a short vertical length scale. Because of the latter, they commonly dominate vertical water-flow differences or 'shear' over vertical length-scales O(10 m), even in areas where tidal motions are dominant (van Haren, 2007). While (surface) tidal motions are highly deterministic, showing a sharp narrow-band peak in frequency spectra from yearlong time series records, (internal) inertial motions are commonly spread over a broader bandwidth of typically 10% of the central frequency, which is often shifted to slightly super-inertial frequencies (Fu, 1981).

The 10%-bandwidth and blue-shift to super-inertial frequencies are partially explained because storms have finite latitudinal extent O(100 km) and near-inertial internal waves can only freely propagate equatorward, to lower latitudes (Munk, 1980). Another explanation is that the ocean contains a considerable variety of rotational vortex-motions or eddies, varying on scales from large O(1000 km), via meso O(100 km) to sub-meso O(1-10 km) and having life-times between weeks and many months.

These eddies provide additional rotational motion or vorticity $\zeta$ to the planetary vorticity f, so that an effective Coriolis parameter results: $f_{eff} = f + \zeta$. The ratio $|\zeta/f| \approx 0.1$ in the open ocean, with a tendency towards blueshift (Perkins, 1976; Fu, 1981). The sign and magnitude of $\zeta$ depend on the sense and sum of vortex-rotation of eddies passing. The associated frequency shift of a near-inertial peak is not attributable to Doppler shift for Eulerian observations via moored instrumentation (Gerkema et al., 2013).

In this paper, an unusual record from a moored pressure sensor is presented that contains sharp, highly deterministic peaks at sub-harmonics of a near-inertial frequency. The record is compared with that from other pressure instrumentation in the same mooring and with those from other variables.

## II. INSTRUMENTATION

A 1275-m long instrumented mooring was located at $\varphi = 37° 00′N$, longitude $\lambda = 013° 44.5′W$, water-depth H = 2380 m on an eastern flank of Mount Josephine (Fig. 1), in the NE-Atlantic Ocean 400 km southwest of Lisbon (Portugal) for 4 months. The mooring consisted of several sections of nylon-coated D = 0.0063 m outer diameter steel cable and an inline-frame approximately halfway the cable. A 4.2-kN elliptic sub-surface buoy was at 1105 m, 100 m above a similar second buoy holding a first downward looking 75 kHz, 20° slant angle to the vertical RDI/Teledyne Longranger acoustic Doppler current profiler 'ADCP'. The buoys were attached via swivels to prevent distortion of the cable due to the rapid rotation of the buoys, at typical rates between 0.2 and 0.06 $s^{-1}$ (van Haren, 2010). A swivel was also mounted between the cable and the acoustic releases just above the 8-kN anchor-weight. The inline-frame held a second 75 kHz downward looking ADCP to cover the lower 600 m range of the T-sensors. This 0.95-kN weighing instrumented frame was not swiveled because of risk of damaging an electric wire it supported. The ADCPs sampled at a rate of once per 900 s water-flow components in 3 Cartesian directions [u, v, w] in 60 vertical bins of 10 m, instrument tilt, temperature T and pressure p.

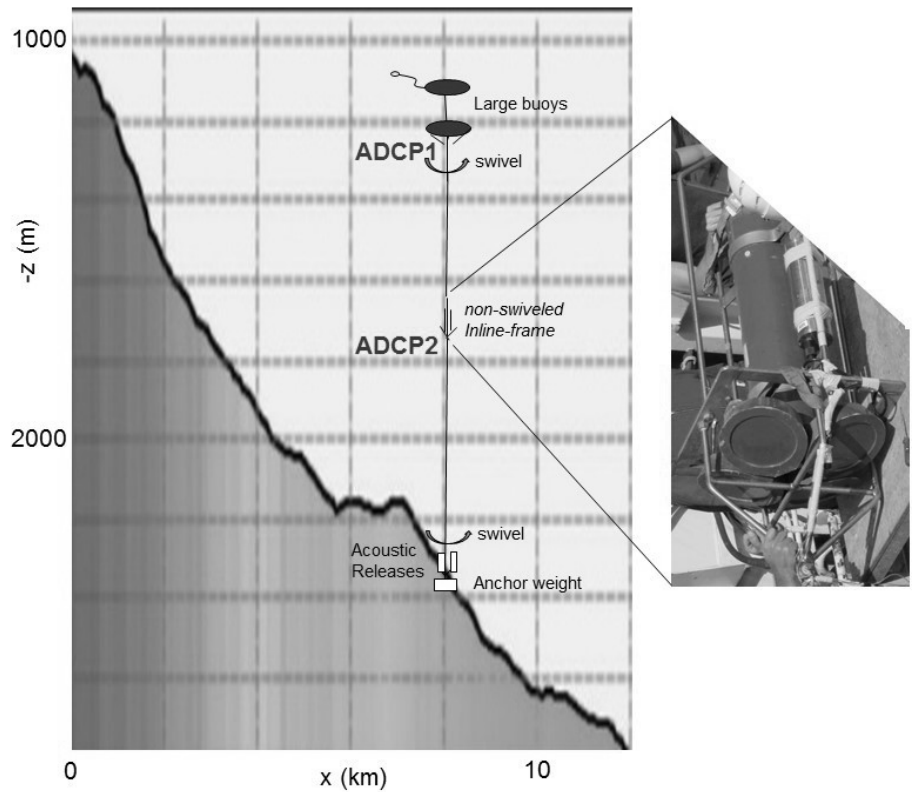

**Figure 1.** Mooring outline sketch over partial Mount Josephine bathymetry from east-west Multibeam transect along 37 °00´N. The bathymetry aspect ratio vertical:horizontal is 8:1. The photo shows the inline-frame with the deep-water ADCP.

The mooring assembly was held taut upright with net 5-kN buoyancy and a net-anchoring of approximately 3 kN. The ADCP's tilt sensors demonstrated that even under maximum 0.3 m s$^{-1}$ current-flow speeds the top of the mooring deflected little, vertically by <0.2 m and horizontally by <20 m, corresponding with a tilt of <1°. The static tension in the cable varied between about 7.5 kN just under upper ADCP1 and 6 kN just under mid-cable ADCP2. The 1.5-kN difference in tension implies a difference of about 0.6 m over 600 m cable length due to extra or reduced stretching for a given cable-section. A twisted-strand steel cable is not rigid and demonstrates some elastic capacities under tension. When vibrating, the inline-frame may thus be considered as a bob between two 600 m long springs, in terms of a deep-sea version of a Wheatstone-device. The difference with Wheatstone's table-top version is that the mooring's top-buoys are not fixed in space like its anchor-weight but can move mainly with horizontal water-flow direction.

Three major oscillations are recognized of the present mooring system. Typical water-flow speeds of U = 0.2 m s$^{-1}$ will cause vibrations in the tensioned taut mooring cable due to vortex shedding at the Strouhal frequency $\omega_s$ = StU/D = 6 s$^{-1}$ (Strouhal, 1878). Here, Strouhal number St = 0.2 is common for open-ocean Reynolds numbers Re between $10^4$ < Re < $10^6$. The inline-frame with ADCP2 has a diameter of approximately D ≈ 0.25 m and will cause vibrations at $\omega_s$ ≈ 0.16 s$^{-1}$. These vibrations may transfer to tension in the cable via rotation, as the inline-frame is non-swiveled. Although it is unclear at what rate the inline-frame rotates, swiveled inline current meters in previous moorings typically swing at speeds of 1-5° s$^{-1}$. The two orders of magnitude slower rotations of the top-buoys are unlikely transmitted in the cable, as they are separated by swivels. Unfortunately, all these vibrations and rotations are not directly measured by the ADCPs because of their relatively slow sampling rate of once per 900 s.

The deep-sea water-flow varies at many frequencies, of which the semidiurnal lunar tidal M$_2$ is the dominant one, at 1.405×10$^{-4}$ s$^{-1}$, and vertical water-flow differences over scales of 50 m are dominant at f ≈ 9.5×10$^{-5}$ s$^{-1}$, as will be shown below. The mooring line thus experiences vibrations or oscillations with periods that vary over four orders of magnitude. Some of the variations in motions are well-predictable, 'deterministic', like those for tidal flow and to a certain extent Strouhal vibrations, although the latter vary their central frequency depending on flow speed. Others are more erratic, intermittent and poorly predictable, like those associated with internal waves and turbulence. The slopes of Mount Josephine are known for the vigorous breaking of internal tides in the lower few 100 meters above the seafloor (van Haren et al., 2015).

## III. OBSERVATIONS

The four-month mean spectra of various parameters from ADCP-sensors demonstrate the common dominant variance peaking at semidiurnal frequencies D$_2$ (Fig. 2a). Some parameters show a smaller peak at, or just higher than, f. This small near-inertial peak is even dominant for relatively small 50-m-scale vertical shear **S** = [du/dz, dv/dz]. Bandpass-filtered near-inertial shear-magnitudes are |**S**$_f$| < 8×10$^{-4}$ s$^{-1}$, which differences are generated by water-flow velocities O(0.01 m s$^{-1}$). This spectral shear-peak

confirms the relatively small vertical length-scale of internal near-inertial motions (LeBlond and Mysak, 1978). Pressure is the only parameter that shows a relatively large peak at diurnal frequencies $D_1$. With temperature T-variations, pressure also shows a peak at quarter-diurnal higher tidal harmonic frequencies. The water-flow ADCP-data are too noisy to distinguish such higher frequency motions.

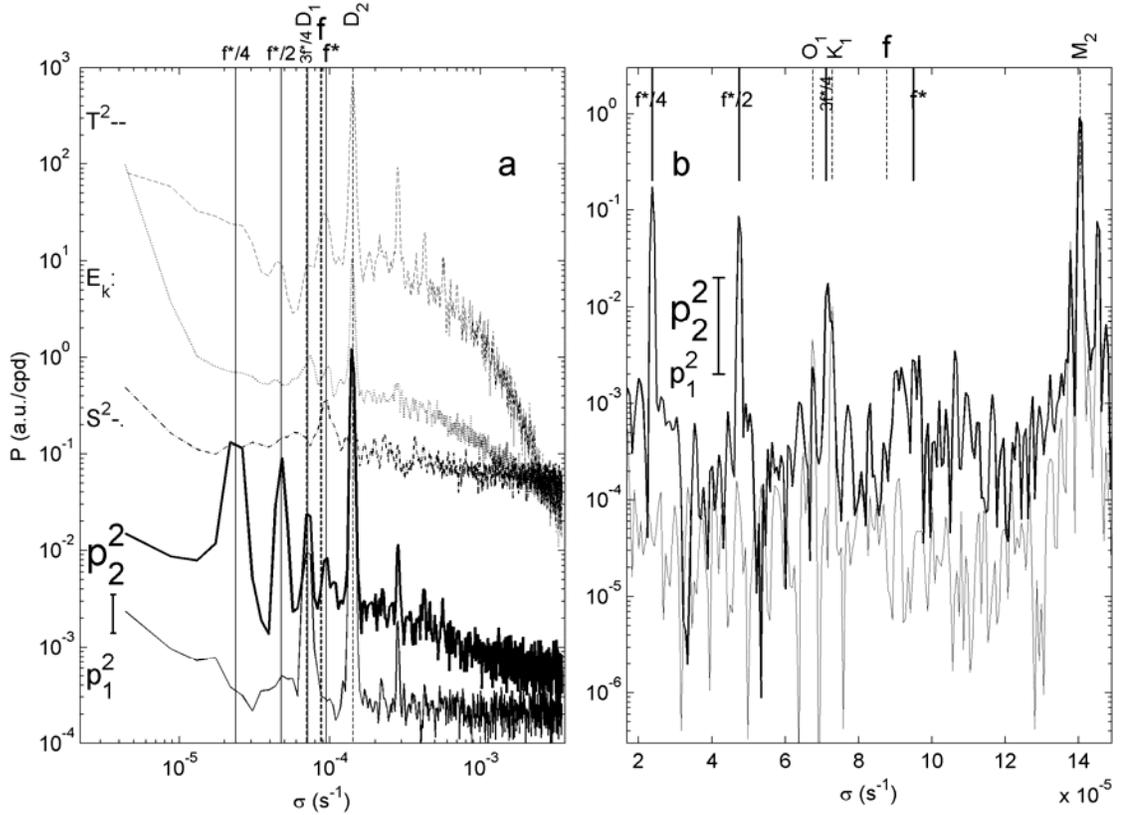

**Figure 2.** Four months mean spectra for upper (thin-solid) and mid-cable (thick-solid) pressure sensor data. For reference, several other mid-cable parameters are plotted of 50-m vertical shear (dash-dotted), kinetic energy (dotted) and temperature (dashed). (a) Moderate smoothing is applied, with approximately 20 degrees of freedom 'dof'. Specific frequencies are described in the text. (b) Magnification of near-raw pressure spectra with 5 dof. The x-axis is linear to preserve the fundamental bandwidth size. Besides the $M_2$-peak, also peaks at solar $S_2$ and $N_2$ are discernible in the semidiurnal band.

An unusual, non-oceanographic sequence of peaks is observed in the spectrum of mid-cable pressure variance $p_2^2$. In comparison with the upper pressure spectrum $p_1^2$, which shows a $D_2$-peak that is indistinguishable from $p_2^2$, peaks appear in the mid-cable spectrum at near-inertial frequency $f^* = 1.083f$, a larger one at $D_1$, and especially at sub-harmonics $f^*/2$ and $f^*/4$, the largest peak of the sequence. Outside the sharp sub-harmonic near-inertial peaks, $p_2^2$ also shows more variance distributed over broader frequency ranges, in comparison with $p_1^2$. This broader range variance is associated with the different mounting of the two ADCPs. The ADCP in a swiveled heavy buoy is less prone to vertical (pressure) excursions than the ADCP in a non-swiveled mid-cable weighted frame. Outside the tidal peaks, the $p_1^2$ spectrum basically represents instrumental white noise as its level is horizontal with almost zero average slope. The larger level of variance in $p_2^2$ reflects 1-m vertical motions, also short-term near the Nyquist frequency. (Here and henceforth, vertical excursions are used as a proxy for pressure variations. In water, a vertical variations of 1 m is nearly equivalent to a pressure variation of $10^4$ N m$^{-2}$.)

Although sharp sub-harmonic near-inertial peaks are only found large in $p_2^2$, there are small elevations in variance around these frequencies in the mid-cable T-spectrum. However, these small elevations are distributed over broader frequency bands like internal tides, instead of the sharp peaks in $p_2^2$ representing quasi-deterministic narrow-band signals like surface tides. Kinetic energy and shear do not show elevated variance around the sub-harmonic near-inertial frequencies.

Closer inspection of nearly raw pressure spectra (Fig. 2b) demonstrates that (surface) diurnal tidal frequencies $O_1$ and $K_1$ (Schuremann, 1941) show distinguishable narrow-band peaks in $p_1^2$, besides the three peaks at main semidiurnal frequencies. However, the dominant diurnal peak in $p_2^2$ is at $\sigma = 7.13 \times 10^{-5}$ s$^{-1}$ = f*/4+f*/2 = 3f*/4, which suggests interaction between motions at the two main sub-harmonics. Comparison with the peaks at highly deterministic diurnal and semidiurnal frequencies demonstrates that the peaks in $p_2^2$ at f*-sub-harmonics are as narrowband. This also holds for the diurnal band, in which $K_1$ is close to, one effective fundamental bandwidth from, 3f*/4. The f*-sub-harmonics mid-cable pressure variations are thus not associated with oceanographic internal wave motions, because they are predictably persistent throughout the entire 4-month record.

In time series that are low-pass filtered with cut-off at $5.8 \times 10^{-5}$ s$^{-1}$ < $D_1$ < f, hence 'sub-inertial' daily-averaged, to remove noise, turbulence and internal waves, the distinction between the two pressure records is obvious (Fig. 3). The $p_1$ shows much smaller amplitudes and a slow trend over the four month record compared with $p_2$. This trend is either due to a settling of the piezo-element of the pressure sensor or of the strands of the mooring cable, as values decrease with time implying that the sensor gets higher in the water column. There is no particular association with T, which decreases over the first month and then increases in the second half of the record. However, this trend in T is also visible in $p_2$. Besides the slow, monthly variation, the T-record also shows a small variation with a periodicity of about 3 days.

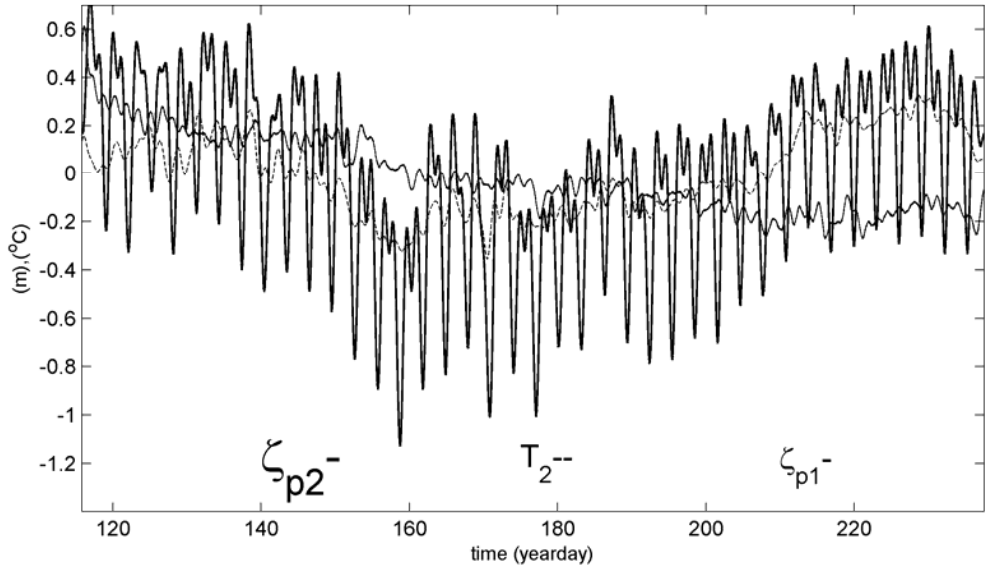

**Figure 3.** Four months record of diurnally filtered data from upper (thin-solid) and mid-cable (thick-solid) pressure and temperature (dashed). All data are plotted with respect to their mean value.

The 3-day periodicity is clear in $p_2$. It reflects the large peak in $p_2^2$ at $f^*/4$. Temperature and pressure move in- and out-of-phase, which indicates the more intermittent character of the sub-inertial T-data compared with the highly deterministic sub-inertial $p_2$-data. The $p_2$ shows every 3 days a sharp downward dip, followed by a longer duration higher level with a smaller amplitude secondary wave. The secondary wave reflects the spectral peak at $f^*/2$. Per 3-day period, the equivalent vertical excursion in $p_2$ between crest and trough amounts about 1.0 m, i.e. an amplitude of 0.5 m.

The 3-day T-variations are about 0.1°C between crest and trough and more variable, whilst lacking a strong one-sided (downward) peak. This implies no direct link between this T-variation and a physical vertical motion of 1 m going upward, which corresponds with a temperature increase of $2\times10^{-3}$°C under the local thermal stratification. It is also not related with a direct pressure effect on the T-measurement, as 1-m upward motion implies a temperature decrease of $-1.5\times10^{-4}$°C following the local adiabatic lapse rate. The observed 3-day T-variation of 0.1°C corresponds with oceanographic >10 m vertical isotherm internal wave excursions, which are much larger than a physical 1-m excursion of the mid-cable ADCP. The pressure and temperature records are thus only indirectly related.

## IV. CONCEPTUAL MODEL OF A DEEP-SEA FOUCAULT-WHEATSTONE DEVICE

The oceanographically anomalous observations have been made using a pressure sensor that was mounted halfway elastic steel cables on a deep-sea mooring. The deterministic-as-tides pressure variations at fractions of the local inertial frequency thus demand a physical explanation in terms of the mechanical set-up. Given that the observations are only made in pressure implies that motions must have a vertical component, with unknown horizontal components. In comparison with the upper pressure sensor in the swiveled top-buoy that is rotationally decoupled from the cable, the mid-cable pressure sensor shows higher variance throughout its record. This implies more vertical motions, possibly also at the Strouhal frequency which is unresolved by the sampling. The mooring cable is less moved by water-flow at inertial periodicity compared to tidal and sub-inertial periodicity, but may feel some effects of the vertical water-flow difference as the 50-m shear peaks at f*. In non-rotating equilibrium, the tension in the cables just above and below the inline-frame changes by about 1 kN.

The exchange of angular momentum between the rotating Earth and the vibrating inline-frame is then envisioned as follows. A Foucault-Wheatstone device of a weight is held between two differently tensioned spring-cables. The tension in the springs acts as restoring force like gravity in a pendulum. Vibrations or rapid oscillations are generated via drag-force by water-flow passing the inline-frame. Compared with the Wheatstone (1854) device (Teunissen, 2022), the deep-sea mooring equivalent is only fixed at the seafloor, while the top may move mainly horizontally. To the knowledge of the author, thus far all theoretical treatment of Foucault-Wheatstone devices have been given for horizontal displacement of a rapidly vibrating bob. None exist on such motions having a vertical component for which, naturally, the direct Coriolis-effect is negligible.

With respect to the equilibrium, the rotating Earth will displace the inline-frame outward, which affects the tension in the cables by generating more via stretching the lower cable to provide the required centripetal force. The tension-imbalance is brought to a new equilibrium by lowering the inline-frame: $p_2$ increases. Strouhal vibrations will act in the direction of the water-flow but by moving back-and-forth they will impose a vertical component in addition to the horizontal one by likewise increasing/reducing the tension, in the lower, fixed cable initially. However, this system unlikely shows a Foucault-periodicity in pressure, as it is symmetric during its rotational motion of the path of vibrations.

An asymmetric situation exists when the mooring is displaced by a steady flow and leaning towards one side away from the vertical, even though tilting very little as in the present case (<1°). When the phase of the vibrational path deflected by the rotating Earth is in the same direction as that of the steady flow, the inline-frame is displaced downward, as indicated above. However, half an inertial period later the inline-frame goes up due to a reduction of tension in the lower cable. This could provide a motion at f*/2. A modulation is expected when the water-flow shear displaces the mooring cable at f*.

When the bob is vibrating with a vertical component as in a z, x/y plane, the common Foucault effect of horizontal motions is directly transferable to the vertical component. The only way to achieve this is via a tilting of the x/y plane. Such a tilt can, e.g., be brought about by shear. As it happens, the 50-m shear peaks at f*, but it is unknown whether this results in sufficient driving of the observed $p_2$ sub-inertial motions.

Acknowledging that solutions are independent of the bob's mass and vibration amplitude, provided that $\omega_s \gg \Omega$, and that the bulky form of the mass is needed to set the vibration in motion due to Strouhal drag-effects, a theoretical model explanation is sought. Such an explanation may be inspired by the presently unknown quarter-inertial period oscillation of the North-pole projection on its equator of a spherical oscillator that is placed on a turn-table (Flückiger et al., 2020).

**V. CONCLUDING REMARKS**

The oceanographically anomalous observations have been made using a pressure sensor that was suspended in the middle of an 1170-m long elastic cable under considerable tension. The vertical excursions of the nearly 1 kN weighing inline-frame are about 1 m crest-trough and have been observed at sub-inertial frequencies. These excursions can thus not be related to freely propagating internal waves that range between [f, N], where buoyancy frequency N >> f (LeBlond and Mysak, 1978). As the pressure variations are found to be highly deterministic at sub-harmonics of f*, they are distinctly different from $M_2$-f = $5.28 \times 10^{-5}$ s$^{-1}$ or $M_2$-f* = $4.55 \times 10^{-5}$ s$^{-1}$. Thus, they do not represent common strongly nonlinear interactions between semidiurnal tidal and (near-)inertial motions. This is also because the observations are not found in water-flow data. The lack of correspondence with known oceanographic internal wave properties thus points at other, mechanical sources.

The primary question is whether f* = $f_{eff}$, and, subsequently, what causes a rather deterministic quasi-permanent background relative vorticity in an oceanographic context. It may be associated with the seamount. Steady and persistent, presumably mesoscale eddies associated with the seamount may be plausible after being generated by the flow around the (large) seamount. However, such persistent eddies have not been observed often in the deep-sea.

Technically, the vertical excursions that are (almost) exclusively registered by the mid-cable pressure sensor are considered to be caused by elasticity following tension-variations in the cable. Around mid-cable, high-frequency vibrations cause vertical excursions of 1 m that are not observed in the upper pressure sensor and which yields the observed much noisier mid-cable pressure signal. Together with the vertical excursions in the direction of the cable, vibrations are expected in the horizontal plane. These horizontal plane vibrations may start rotating with the Earth like a Foucault pendulum. But apparently their motions are not fixed in the horizontal plane, and we are left wondering what causes the near-vertical increase and relation of tension at mid-cable. Oceanographically, a pressure sensor normally does not register quasi-horizontal inertial motions, cf. the $p_1$-record.

The mid-cable pressure observations show a resonant behavior at sub-inertial frequencies that is reminiscent of a Foucault-Wheatstone device, albeit for the first sub-harmonic only. The observed peak at the second sub-harmonic is more puzzling, as has been reported for a spherical oscillator in a rotating frame of reference (Flückiger et al., 2020). The present deep-sea mooring differs from common Foucault-Wheatstone devices because it is not fixed at the top as the buoys may move freely, although held under considerable tension. Wheatstone (1854) mentions that whenever 'the wire be caused to vibrate in any given plane', we may also hypothetically view vibration in the direction of the cable, so near-vertical. Such vibration would cause large pressure vibrations as observed, possibly a unique case of mooring cable resonance. However, it is not expected that vertical excursions are not accompanied by horizontal vibrations, so that 2D-motions result. Perhaps there is an analogue with a spherical oscillator. This requires future studies.

**Acknowledgments**

I thank captain and crew of R/V Pelagia for their assistance and technicians of NIOZ-NMF for help in preparing and deploying the mooring.

**AUTHOR DECLARATIONS**

**Conflict of Interest**

The author has no conflicts to disclose.

**DATA AVAILABILITY**

The data that support the findings of this study are available from the corresponding author upon reasonable request.